安远 AI
CONCORDIA AI

智谱

# Frontier Open-Weight AI Risk Management Framework

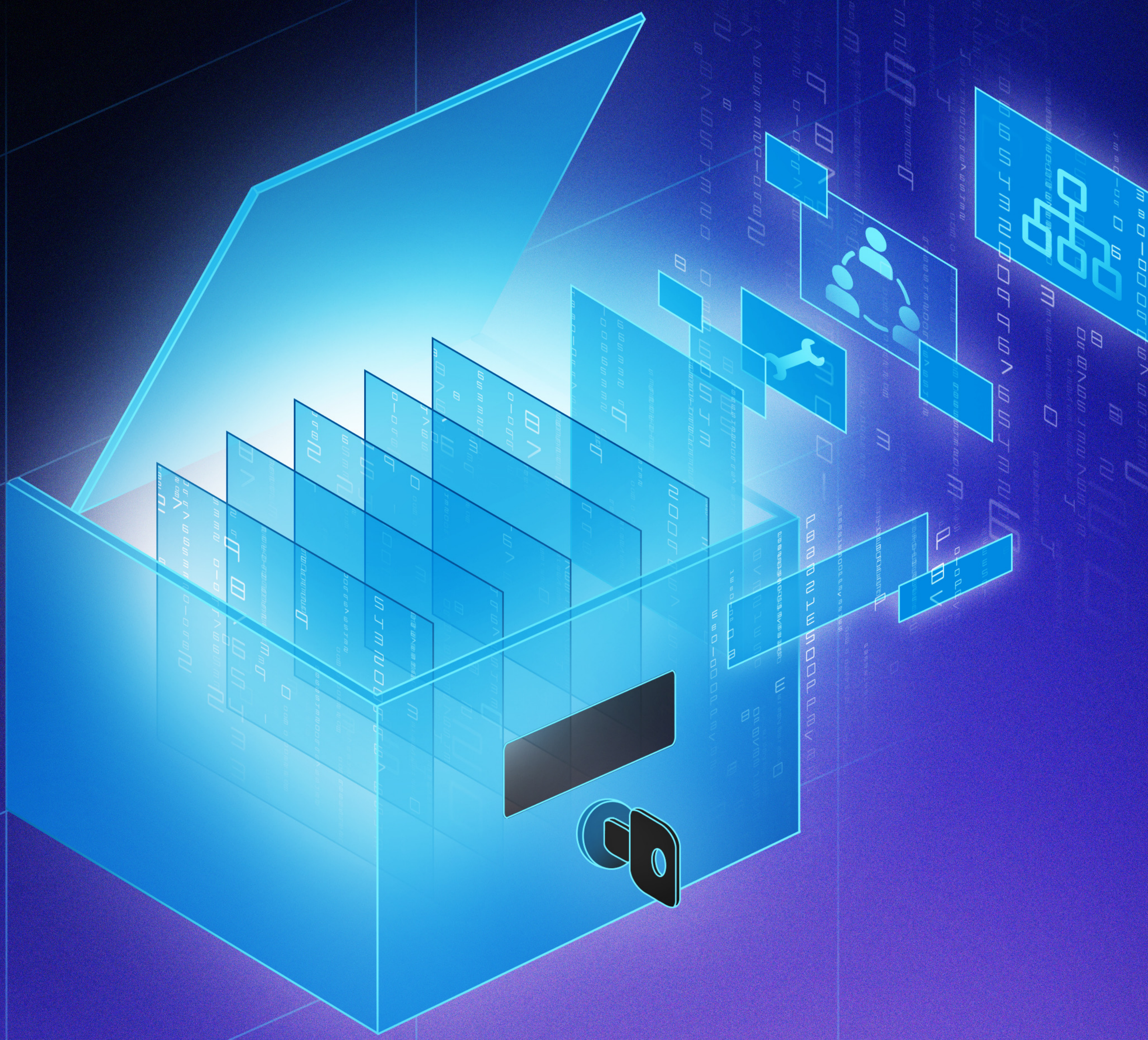

September 2026

# Executive Summary

Open source became the foundation of modern software. Open weights are becoming the foundation of global AI, shaping how frontier AI reaches the world: models, tools, and harnesses that any organization, in any country, can inspect, adapt, and deploy on its own terms. That reach brings both significant benefits and unique governance challenges. Proprietary systems offer controlled deployment and centralized oversight; open-weight systems accelerate academic research, drive widespread economic innovation, increase transparency, allow data to remain under local governance, and complement closed systems with customizable defenses that reduce dependence on any single point of failure.

Yet openness carries distinct, irreversible risks. Open-weight models can be fine-tuned to strip away safeguards, repurposed for malicious ends, and distributed beyond any central point that can take accountability. The major concerns are dual-use capability and unmonitored misuse. Major reports of the scientific consensus on AI, including the *Preliminary Report* of the United Nations Independent International Scientific Panel on AI, the *International AI Safety Report 2026*, and the *Singapore Consensus on Global AI Safety Research Priorities*, all emphasize the urgency of addressing these risks, without undermining the societal benefits of open innovation.

This framework addresses the governance challenge at the heart of that choice. Co-developed by Concordia AI and Z.AI, it adapts the full risk management lifecycle—from risk assessment to mitigation and governance—to the unique properties of open-weight models. It serves as a companion to the Frontier AI Risk Management Framework 2.0 co-developed by Shanghai AI Lab and Concordia AI.

This document does not resolve every tension between openness and safety. Rather, it provides the first comprehensive, evidence-based foundation for navigating these tensions. As frontier open-weight releases accelerate, risk management best practice and standards are urgently needed. We invite researchers, developers, policymakers, and civil society to build on this framework and advance the science and practice of open-weight AI governance.

# Contributions and Acknowledgements

**Contributors**
Concordia AI: Brian Tse, Oskar Galeev, Weibing Wang, Yuan Cheng, Liang Fang
Z.AI: Wenbin Qiao, Xiaochen Wang, Chen Zhang

**About the organizations**
Concordia AI: Concordia AI is a mission-driven, independent social enterprise focusing on AI safety and governance. Concordia AI aims to steer AI development to harness transformative benefits while identifying, mitigating, and coordinating global preparedness for catastrophic risks from frontier AI.

Z.AI: Derived from technology transfer of Tsinghua University, Z.AI integrates its AGI mission with the aspiration for domestically-developed large-scale models. It is China's first and one of the world's early enterprises dedicated to developing AGI large models. Since its inception, the company has successively rolled out China's first 10B-parameter model, first open-source trillion-parameter model, first conversational model, first multimodal model, as well as the world's first device-controlling Agent.

**Acknowledgements**
We thank the participants for their valuable inputs and discussions at the Open-weight AI Risk Management Workshop, co-hosted by Concordia AI at the International Association for Safe and Ethical AI 2026 in Paris, and at the Open-weight AI Governance Workshop, co-hosted by Concordia AI and FAR.AI at the World AI Conference 2026 in Shanghai.

We are also grateful to Stephen Casper, Mick Yang, Jingwei Yi, Chaozhuo Li, Gabriel Wagner, and Kwan Yee Ng for providing feedback and suggestions on this framework.

**How to cite this report**
Concordia AI and Z.AI. 2026. Frontier Open-Weight AI Risk Management Framework.

# Table of Contents

# Framework Overview

The *Frontier Open-Weight AI Risk Management Framework* is designed to address the unique and complex risks associated with widely available open-weight general-purpose AI models. Developed as a complement to the Shanghai AI Lab and Concordia AI's *Frontier AI Risk Management Framework 2.0* [1] (hereinafter referred to as *Framework 2.0*), it is tailored to empower developers and the broader open-weight ecosystem with actionable protocols for proactive risk management. By establishing technical safety guidelines for the open source community, it also aligns with the evidence-based assessment of opportunities, risks and impacts of AI recommended by the United Nations Independent International Scientific Panel on AI [2]. The UN Panel, together with other major scientific consensus efforts [3, 4], emphasizes the urgent need to manage dual-use capability risks. In addition, leading national standards bodies such as TC260 recommend closer collaboration between open-source model developers and open-source communities to strengthen rules and responsibilities for risk disclosure, clearly defined prohibited uses, and clarified security obligations to prevent misuse [5].

This framework is structured around a six-stage lifecycle: risk identification, risk thresholds, risk analysis, risk evaluation, risk mitigation, and risk governance. This six-stage structure mirrors the set of protocols laid out in the *Framework 2.0*. At each stage, this framework applies the corresponding risk management stages to open-weight release. Many protocols in the *Framework 2.0* assume that the developer retains control over deployment, for example the ability to gate access, monitor use, or roll back a release. Open-weight release fundamentally decentralizes these control levers, moving the capacity for oversight and mitigation from the original developer to the downstream ecosystem. The framework identifies which protocols rely on these developer-side control measures, flags where this reliance breaks the protocol for open-weight releases, and specifies an adapted approach that does not depend on developer-side control.

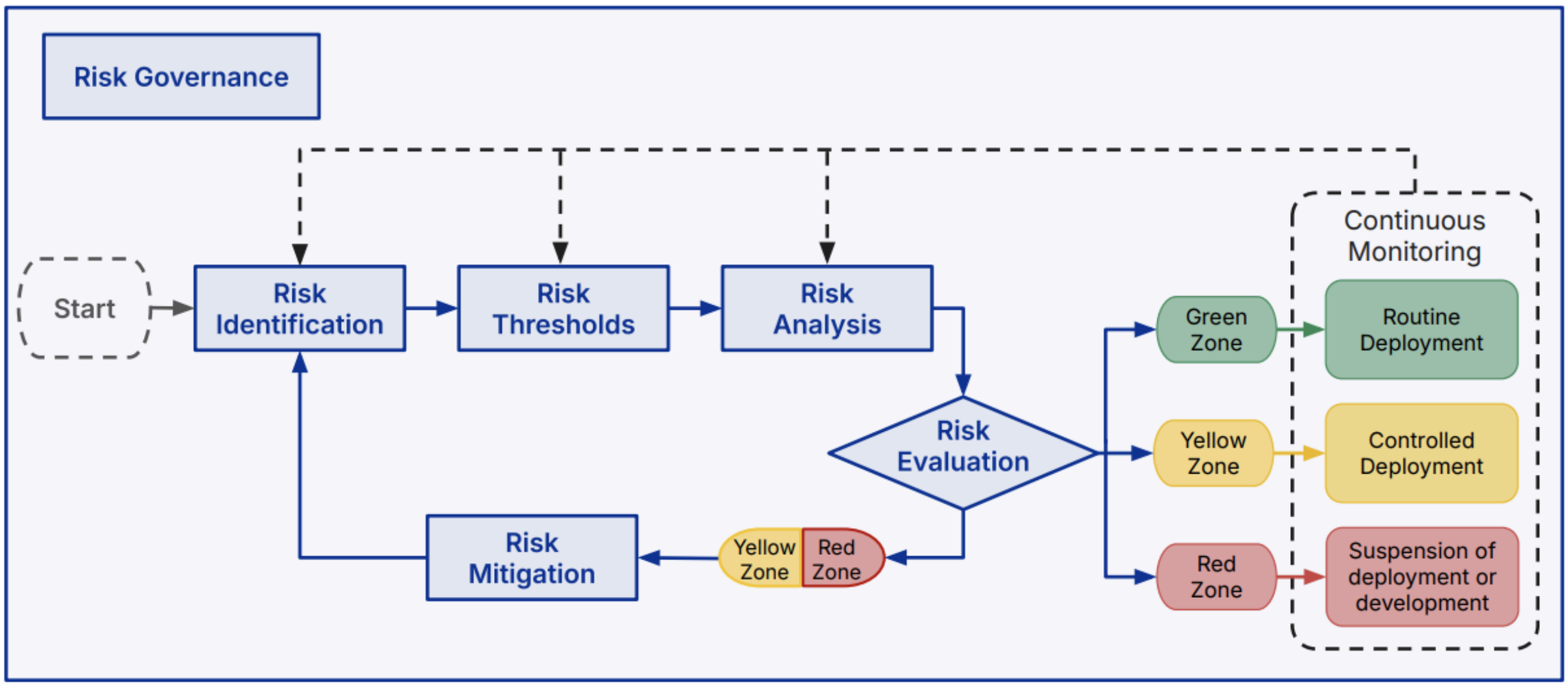

This stage-by-stage adaptation is necessary because, while open-weight models offer substantial benefits, such as accelerating innovation, expanding access, and fostering robust external safety evaluations, these very same features introduce distinct governance challenge the original developers cannot control and monitor in downstream use; safeguards can be removed more easily; and malicious actors can fine-tune open-weight models for harmful purposes. This document explores these challenges and the tradeoffs they involve. It does not offer definitive answers, but works toward understanding how the benefits of openness might be realized without compromising public safety and security.

## Definition & Scope of Frontier Open-Weight General-Purpose AI

This framework applies a frontier AI model risk management approach to a narrow subset of open-weight general-purpose AI models: those whose capabilities in risk-relevant domains approach or match the current frontier. The definitions and governance mechanisms for open weights are still nascent, and there is no consensus about what risk management approaches are needed [6]. It is therefore important to precisely define this scope, because governance attention should focus on instances where openly releasing weights has the most consequential risks and benefits [7], without imposing frictions on the broader open ecosystem.

An open-weight model is one whose parameters are publicly available for download. While publicly released frontier models are often colloquially called “open source”, most of these models should be characterized as “open-weight” as the release does not include the training pipeline or datasets [3] (i.e. they are not fully open). Releasing weights alone is sufficient to meet the definition of an “open release” [8].

While there is no single agreed definition of AI openness, recognized approaches include the Open Source Initiative's Open Source AI Definition [9] and the Linux Foundation's Model Openness Framework [10], both of which treat openness as a spectrum defined by what model components are released and under what licenses. This framework discusses the risks and benefits of openness in this sense, as a spectrum rather than a binary property.

Open-weight releases are one-directional [4, 11]. Any proprietary models can subsequently be opened, but open-weight models cannot be retroactively closed. This asymmetry is one of the defining features of open-weight risk management.

# Benefits of Open-Weight General-Purpose AI

Risk management of open-weight models should start with a recognition that AI openness provides substantial benefits, and that these benefits depend on how much is shared (whether only the model's weights are open, or also its training data, training code, and documentation). As the *International AI Safety Report 2026* states, these benefits may outweigh the additional risk of releasing weights openly [3]. This framework addresses the case in which that tradeoff is most consequential for a wide number of downstream deployers: frontier open-weight models.

The benefits of open-weight release fall into four broad categories. The first is **transparency and oversight**: openness can enable external actors to scrutinize and verify properties of frontier AI models. The second is **facilitating AI research,** especially among AI safety organizations and academic institutions. The third benefit is **wider adoption and innovation** through expanded and democratized access to the model. The fourth category includes **benefit-sharing and diffusion of power**: open weights distribute the capacity to develop, evaluate, govern, and benefit from AI beyond a small set of powerful countries and labs.

## Transparency and Oversight

Open-weight models are subject to external accountability in a way that closed-weight models are not. Transparency benefits include:

- **Independent scrutiny:** researchers outside the developer can audit a model and its accompanying components for bugs, biases, and discrepancies between developer claims and actual behavior [12, 13].
- **Broader evaluator pool:** open release directly extends oversight over frontier models to civil society, regulators, downstream developers, and the research community. This broader pool can not only identify critical safety flaws, but also set best practices for beneficial applications [3, 13].
- **Realistic red-teaming:** red teams can simulate adversarial attacks using the same publicly available tooling that adversaries might exploit, enabling more realistic emulation of attacks than black-box probing allows [12].

## Facilitating AI research

Open weights lower the barriers for frontier AI and AI safety research. They enable:

- **Alignment and interpretability research:** some interpretability research and frontier alignment research requires direct access to model parameters and patterns, which API interfaces alone do not provide [12].
- **Scientific reproducibility:** open release enables independent peer review and replication of frontier AI work, allowing to verify developer claims about capabilities and safety properties [8, 14]. We acknowledge that open weights alone do not allow for full end-to-end training reproducibility without the underlying datasets and training code.

- **Cumulative progress:** open release allows less well-resourced researchers to build upon existing systems, supporting cumulative scientific progress [3].

## Adoption and Innovation

Open weights expand the contexts in which AI can be usefully deployed. They improve:

- **Defensive capabilities:** defenders can inspect, adapt, and run open-weight models on their own infrastructure for vulnerability research and incident response, improving response speed and reducing dependence on a small number of closed providers [15, 16, 17]
- **Downstream development efficiency:** open weights allow developers to build on top of frontier capabilities, reducing the compute, capital, and time required to develop new applications [12, 18].
- **Privacy and offline deployment:** open weights enable on-device and local deployment for cases where sensitive data must remain on specific premises, or where reliable internet connectivity is unavailable [3, 12].
- **Localization for underserved users:** open weights make it easier to fine-tune models for low-resource languages, regional contexts, and specialized professional tasks that proprietary developers might not prioritize [3].

## Diffusion of Power and Benefit-sharing

Open weights distribute the capacity to develop, govern, and benefit from AI beyond a small set of incumbents.

- **Market deconcentration:** open weights lower barriers to entry, foster competition, and broaden the developer community, countering the concentration of AI capability in a small number of well-resourced firms [12, 15].
- **Bridging the digital divide:** open release allows communities in developing regions to access, study, and adapt frontier AI, addressing the digital divide and enabling global participation in AI development that would otherwise be foreclosed [3, 19, 20, 21].
- **Decentralized governance:** open weight releases expand the set of actors with meaningful influence over how AI is developed, evaluated, and governed, supporting decentralized mitigation strategies and oversight that reflect perspectives beyond those of frontier labs [13].

These benefits of openness explain why open-weight release matters and holds so much potential, including for improving AI safety and security. However, they do not by themselves determine whether and how any specific frontier model's weights should be openly released. That question is the starting point of the risk management lifecycle. The rest of this framework works through the six lifecycle stages, weighing openness against its marginal risks at each stage.

# Risk Identification

Risk identification is the process of finding, recognizing, and describing risks. The IEC/ISO 31010 [22] international standard on risk management covers both evidence-based reviews and systematic expert review, and specifies that irrespective of the method used, “it is important that due recognition is given to human and organizational factors when identifying risk”. For general-purpose AI systems, the *International AI Safety Report 2026* identifies threat modeling and risk taxonomies as the most established identification methods [3].

This section describes how open-weight release amplifies or alters the misuse and accident risks that *Framework 2.0* identifies for frontier general-purpose AI models [1]. It deliberately does not cover loss-of-control or systemic risks, since there is no substantial evidence that open-weight release exacerbates these relative to proprietary release.

Risk identification for frontier open-weight GPAI models is crucial, because these models exhibit several of the contributing characteristics of risk: open-weight release is difficult or impossible to reverse, models reach a wide audience through decentralized distribution, and their capabilities increasingly sit at the frontier. Decentralized release also makes it more complicated to track the real-world spread, usage, and impact of open-weight models, which makes identifying all of their risks more challenging [6]. The list of risks in this version reflects the current state of the evidence. The taxonomy will be updated as new risks emerge, and loss-of-control and systemic risks will be added if future evidence shows that they are amplified by open-weight release.

## Misuse Risks

- **Loss of visibility for developers:** Once weights are openly released, developers lose the ability to observe or trace the model’s ultimate use. They cannot monitor deployments, detect patterns of abuse, or attribute specific instances of misuse back to a source, which makes harmful applications both harder to prevent and harder to remedy after the fact [2, 3]. The risk is that a broad range of actors can circumvent built-in safeguards, such as content filters, and fine-tune models for malicious applications [23]. Without visibility into deployment, developers have no way of knowing when this occurs, thereby increasing the risk of abuse.
- **Lowering the barriers for well-resourced malicious actors:** Open-weight models may enable a wider range of actors to augment their existing capabilities for malicious purposes and without oversight [24]. Where deployment costs are low, the effort required to repurpose open-weight models for harm falls accordingly [25]. However, at the frontier, the barrier is higher: since substantial compute and technical expertise are needed to serve and fine-tune the largest open-weight models, this risk is concentrated among well-resourced actors for frontier open-weight models.
- **Tamperability:** Since open-weight models are customizable, malicious actors can strip out safety features [26]. Recent research shows that such malicious fine-tuning of open-weight models can enable dangerous applications [27, 28, 29]. Once a maliciously fine-tuned version exists, it can be further disseminated, so a single instance of modification can put a model with compromised safeguards in the hands of a large number of users.

- **Absence of enforceable AI content labeling:** Open-weight release fundamentally undermines the ability to ensure that AI-generated content carries reliable provenance markers. Whereas proprietary API providers can embed watermarks, metadata tags, or content credentials at the inference layer, open-weight models can be run locally without any such instrumentation, and any labeling mechanisms built into the model weights can be stripped or circumvented through fine-tuning, weight editing, or inference-time modification [30]. This directly amplifies misuse risks: malicious actors can generate large volumes of synthetic text, images, audio, or video without any detectable signature, making it substantially harder for platforms, fact-checkers, and end users to distinguish AI-generated disinformation, deepfakes, and phishing materials from authentic content. In turn, this lowers the cost and increases the reach of large-scale persuasion and harmful manipulation campaigns, while also complicating post-incident attribution and regulatory enforcement.

## Accident Risks

- **Unintended degradation of safety mechanisms:** Aligned models' safety guardrails can be eroded during fine-tuning even without malicious intent: benign and commonly used datasets have been shown to inadvertently degrade safety alignment when used in fine-tuning. A model's initial alignment is therefore not necessarily preserved after custom fine-tuning, introducing new safety risks that current infrastructures fall short of addressing [31].
- **Propagation of flaws:** When a small number of open base models sit under a large share of downstream applications, unresolved flaws such as embedded biases, adversarial vulnerabilities, or the ability to game post-deployment monitoring are distributed across shared components, producing correlated and cascading failures [4, 24, 32]. Unlike proprietary models where a host can universally roll out a fix, open-weight updates only work if downstream developers voluntarily adopt them, and there is no guarantee that they will.

# Risk Thresholds

Risk thresholds are clear boundaries for unacceptable risks. They determine decisions about model release and access, as well as mitigations and governance [1, 33]. While the development of risk thresholds has been a frequently stated priority in international AI governance, risk thresholds specific to open-weight models have not yet been articulated [34]. The *UN Preliminary Report* argues that shared and standardized risk thresholds are urgently needed [2]. But openness demands new governance approaches, since many of the controls available for fully closed proprietary models no longer apply once weights are released [35].

This section adopts the red and yellow lines from the *Frontier AI Risk Management Framework 2.0*. "Red lines" are unacceptable outcomes; "yellow lines" are early-warning indicators for escalating safety and security measures. These include domain-specific red line specifications across several critical areas that could threaten public safety and national security, including cyber offense, biological threats, large-scale persuasion and harmful manipulation, and loss-of-control risks. For details, see Tables 2.1 to 2.4 [1]. Because deployment-side mitigations are largely unavailable once weights are released, the margin of safety has to be set differently in the case of open-weight models [36].

To calibrate this open-weight-specific margin of safety, this framework employs four interconnected analytical dimensions that together allow us to estimate the likelihood and severity of potential harm: Deployment Environment, Threat Source, Enabling Capability, and Societal Resilience.

## Deployment Environment (E)

This dimension refers to the operational context and constraints within which the AI model is deployed. Examples include deployment domain, operational parameters, regulatory environment, user demographics, infrastructure dependencies, and available oversight mechanisms. Changes in the deployment environment can significantly alter an AI model's risk profile, even if its capabilities remain the same.

**Since open-weight foundation models are disseminated and adopted in a decentralized way, it is particularly complex to assess risks from their interactions with their deployment environment.** With open-weight models, the operational context can vary widely: anyone can download the weights, fine-tune them, and deploy them across domains like healthcare, finance, or unregulated personal projects, and across different regulatory environments and user demographics. This creates interconnected vulnerabilities, since changes in infrastructure dependencies or oversight mechanisms are harder to track, potentially leading to cascading failures if a modified version is embedded in a critical system. Proprietary models, by contrast, are typically deployed through controlled channels such as APIs, allowing providers to enforce consistent parameters and monitor usage, even though their view of actual downstream use cases remains narrow.

Because open-weight models lack centralized post-deployment controls, the same dual-use capability in an open-weight release carries a higher probability of unmitigated misuse.

Consequently, pre-release capability evaluation triggers need to be set more conservatively. Proprietary models can tolerate higher risk thresholds, since built-in safeguards and restricted access absorb some risk before it escalates (though this assumes effective internal governance). Open-weight models instead shift the burden of risk management downstream to users and communities.

## Threat Source (T)

Threat source is defined as the origin or agent that could trigger harmful outcomes through interactions with the AI model. Examples include external actors (malicious users, adversaries), internal factors (model misalignment, training data biases), operational factors (human error, system integration failures), and emergent behaviors arising from complex AI-environment interactions.

Open-weight models widen the set of threat sources across each of these categories: external actors can more easily gain access to modify and exploit the model; community fine-tuning introduces emergent behaviors that add unpredictability; and operational failures across varied deployment settings are harder to anticipate without centralized control. In principle, proprietary models have more limited threat sources, since providers can monitor usage, rate-limit access, and revoke misused accounts. The contrast is less clean in practice, however: fine-tuning APIs can override safety guardrails even in hosted models, and user-facing interfaces on closed systems have themselves been vectors for misuse [33].

To calibrate a risk threshold against the threat source dimension, we must know which actors can realistically make use of frontier open-weight capabilities. There are at least four relevant types of actor:

- **Ordinary users** do not modify weights, relying on existing inference frameworks, published jailbreaks, and other off-the-shelf bypass tools.
- **Skilled developers** use single-node compute and publicly available LoRA, fine-tuning, and model-editing tools.
- **Professional teams** have the engineering capacity for full fine-tuning, model editing, agent engineering, and system-level optimization.
- **Well-resourced actors** possess technical expertise, proprietary data, as well as financial and compute resources comparable to a developer team.

The interaction between these tiers matters more than any tier in isolation. Low-resource users do not need to perform difficult modifications themselves, since a professional team or well-resourced actor can strip safeguards and re-release an unsafe derivative that ordinary users can then download and run directly.

**A single open-weight release exposes a model to every tier at once, with no mechanism to admit some actors and exclude others,** which is why capability thresholds for open-weight models may need to sit lower than for proprietary equivalents. Proprietary models can tolerate higher capability thresholds by relying on gatekeeping to filter threats, though this may overlook vulnerabilities that only surface in black-box deployments. Open models, on the other hand, could benefit from community-driven threat identification, which may allow capability thresholds to be adjusted upward over time through collective safety efforts.

## Enabling Capability (C)

Enabling capability refers to the core functional abilities of the AI model that make specific risk scenarios possible when the model is deployed without additional safeguards. This includes both intended capabilities such as scientific reasoning, coding, and planning, and emergent capabilities that may arise from scale or training. Particular attention is paid to bottleneck capabilities—those whose presence or absence most directly determines whether a given harmful outcome can be realized.

For open-weight models, enabling capability has to be assessed not against the model as released, but against the capability that a downstream well-resourced actor can extract from it through fine-tuning. Proprietary models, by contrast, can be assessed against the capability of the deployed system as the provider intends it to run, with system prompts, refusal training, and monitoring in place. **Open-weight risk assessment therefore has to track the upper envelope of what the weights make possible (i.e. worst-case scenario).**

If proprietary models cross a certain malicious use threshold when their (as-deployed) capability exceeds a given level, open-weight models should be treated as having crossed the same threshold at a *lower* level of (as-released) capability. This is because Deployment Environment (E) becomes more variable and permissive once weights are public, and Threat Source (T) expands as a wider set of actors can modify the model. In practice, this lowers the bar for what should count as a dangerous capability in an open-weight release: capabilities that would sit safely below the threshold under proprietary deployment can cross it when the deployment environment and potential threat sources are more risky.

## Societal Resilience (R)

Societal resilience is a society's capacity to withstand and recover from AI-induced harms. The assessment of societal resilience should be multifaceted, incorporating factors such as public education programs on misinformation detection (e.g., widespread digital literacy initiatives that teach users critical thinking and fact-checking skills to reduce their vulnerability to manipulation), advancements in cybersecurity infrastructure that shift the offense-defense balance (e.g., through robust protocols, AI-assisted threat detection), and enhanced biosecurity measures (e.g., mandatory DNA sequence checks by providers to prevent AI-enabled bioengineering misuse) [4, 37].

Open-weight releases can strengthen societal resilience (R) in ways that closed deployments cannot. By placing model weights directly in the hands of defenders, open release enables localized defenses, faster responses, crowdsourced safety tools, and reduced dependency on centralized infrastructure. However, whether this successfully lowers net risk depends on whether the pace and reach of these defensive adaptations can outpace the speed at which open-weight capabilities improve, and threat actors gain access to them.

**Taken together, the framework recommends setting risk thresholds through a holistic judgment based on these dimensions in combination, rather than having a test for each dimension. Deployment Environment (E), Threat Source (T), and Enabling Capability (C) set the level of risk exposure that a given open-weight release**

**generates, while Societal Resilience (R) sets the level of risk that a society can safely absorb before that exposure becomes unacceptable.** A threshold is crossed when E, T, and C together outpace what R can absorb, which means that the same as-released capability can sit above or below the risk threshold depending on the resilience of the receiving society at the time of release.

# Risk Analysis

Risk analysis should be conducted throughout the entire AI development lifecycle to determine whether the AI has crossed yellow lines, i.e. displayed the early warning indicators that require safety measures to be escalated. We recommend that AI developers conduct pre-development and pre-deployment analyses to inform critical deployment decisions, as well as continuous post-deployment monitoring to guide the safe development of next-generation systems. The subsections below highlight how open-weight release complicates the risk profile at each stage of analysis. Most importantly, pre-deployment analysis must consider the distinct risk factors of open-weight release, rather than importing evaluation practices designed for closed-weight models [11]. Post-deployment monitoring is constrained by the limited traceability of open-weight models once their versions spread [38].

## Pre-development

Pre-development risk analysis for open-weight release should focus on design and training-time decisions that will shape the model's downstream risk profile once the weights are public and that cannot be revised after release. Two techniques are central at this initial stage of the model development pipeline:

- **Training data curation:** adversarial content in pre-training data or in downstream fine-tuning can compromise safety alignment. Iterative and structured curation of training texts has been shown to reduce models' susceptibility to jailbreaks, even when the developers do not know what techniques and procedures the attack will involve [39].
- **Anticipated safety gap:** Developers should model the difference in dangerous capabilities between a model with intact safeguards and one that has been stripped of them. Evidence shows that this gap widens with model scale, meaning that pre-development choices about target scale and safety training investment have to be based on the anticipated safety gap, not just the as-trained capability (with safeguards) [40].

## Pre-deployment

When conducting pre-deployment risk analysis for open-weight release, developers must reason about a distribution of possible downstream model variants rather than a single deployed system, since downstream actors can fine-tune and modify the model after release. Two broad categories of analysis address this:

### Worst-case risk analysis techniques

- **Simulating model tampering attacks:** Tampering attacks involve modifying a released model directly, whether through malicious fine-tuning, weight edits, or latent activation changes, to elicit maximum capabilities in high-consequence domains. Standard input-output evaluations can only indicate the lower bound of worst-case behavior, whereas tampering simulations provide tighter upper bounds by testing what a downstream actors could actually extract from the weights [41]. For example, in recent research, open-weight models have been tailored for biological and cyber misuse, in order to compare these capabilities against open- and closed-weight baselines and estimate the added risk from this kind of tampering [23].

- **Full-access audits:** Because tampering involves direct modification of the model, audits of open-weight systems require full access to weights and internal states [42]. Worst-case capability evaluations under adversarial fine-tuning should be treated as a necessary step for open-weight release [41].
- **Tamper-resistance as a worst-case evaluation target:** Because open-weight developers cannot rely on system-level safeguards, worst-case analysis has to cover the full range of interventions that a downstream actor could apply to weight-level safeguards, including fine-tuning, steering, model editing, pruning, quantization, distillation, model merging, and backdoor insertion. Current off-the-shelf safeguards typically fail across this range, which makes tamper-resistance one of the central open technical challenges for open-weight model risk management [6, 43].

### Marginal risk analysis techniques

- **Baseline capability comparison:** Open-weight releases should be evaluated against existing tools and proprietary models to isolate their true "marginal risk." If an open model merely replicates capabilities that are already cheap and accessible elsewhere, its net added risk is minimal [7]. While this marginal risk lens helps resolve debate over open-weight safety, most current evaluations still fail to measure models systematically against these baselines [24].
- **Evaluation gap and risk evolution:** Baseline comparison at the point of release does not settle the question of how much harm an open-weight release could cause downstream. The DeepSeek-R1 release illustrates that if a model remains vulnerable to jailbreak, enhanced reasoning makes misuse more practically feasible, and downstream fine-tuning can further compromise safety protections [44]. Pre-deployment safety testing has itself become harder to conduct, as frontier models increasingly distinguish between test settings and real-world deployment and exploit loopholes in evaluations [3]. Both problems can be addressed by explicit delta evaluation triggers. Developers should re-test the affected components whenever a material change occurs in training data, post-training methods, model architecture, tool-use and agent capabilities, or the release mode, rather than defaulting to prior safety assessments.

Marginal risk analysis allows risk mitigation to be calibrated to the release's capability relative to the closed-source frontier. Releases that exceed frontier closed-weight capabilities warrant the most intensive pre-release evaluation, third-party audit, and continuous post-release monitoring. Releases at frontier parity require standardized evaluation protocols and shared safety guidance. Releases well below the frontier only warrant lighter-touch measures such as simplified evaluation and community and industry self-governance.

## Post-deployment

- **Monitoring of open-weight releases is uniquely constrained:** once weights are downloaded, there is no centralized point through which the developer can observe how the model is being used, adapted, or redistributed, which makes downstream usage effectively decentralized, and the model impossible to retract [45, 46]. Some emerging techniques attempt to mitigate this by embedding identifying signals in the released model, either as generation-time watermarks or as fingerprints that allow derivatives to be traced back to their source, though both signals are harder to sustain in the open-weight setting than in the closed-API setting [47, 48]. Even with these emerging techniques, post-deployment monitoring for open-weight models remains methodologically underdeveloped, a gap acknowledged by developers themselves [49].

- **Indirect and proxy monitoring techniques:** Where developers cannot directly observe downstream use, they can draw on indirect signals, such as API abuse on comparable proprietary systems (to infer patterns of misuse in downloaded open-weight variants), technical monitoring of publicly hosted derivatives, and shared risk early-warning and incident-response mechanisms across the ecosystem. These techniques do not close the visibility gap, but they can surface classes of misuse for the risk management process.

# Risk Evaluation

This section outlines an approach that involves classifying models into three zones based on their risk level: green zone (safe to deploy with standard measures), yellow zone (requiring enhanced safeguards and authorization), and red zone (requiring extraordinary measures such as development and deployment restrictions). We recommend iterative risk-reduction measures until risks reach acceptable levels.

As in the general frontier AI risk management process, this zone classification directly determines what mitigation measures and governance protocols are required. The main difference for open-weight releases lies in the yellow zone. A critical challenge in the evaluation of open-weight models is the durability of safeguards, since many of the restricted access and controlled deployment approaches that can typically be used to manage yellow-zone risks become obsolete once weights have been publicly released. Therefore, deployment decisions have to account for this constraint:

- **Residual risk assessment depends on safeguard durability.** Once weights are released, safety training is the only mitigation that continues to apply, so residual risk depends largely on whether that training will survive when the model is further trained after release. Safety training can be undone with a small number of harmful training examples [31]. An attacker does not even need to collect those examples, because the aligned model can be prompted to generate them itself [29]. Current methods for making models refuse harmful requests or forget dangerous knowledge can often be removed with a few steps of fine-tuning [43]. The evaluation methods used to assess these defenses can also mislead reviewers into treating safeguards as more durable than they are [50]. Building durable safeguards and evaluating them rigorously remain among the open technical problems in open-weight model safety [6]. Residual risk estimates for open-weight models therefore carry wider uncertainty than the same estimates for models kept behind a controlled API, and this uncertainty should be reflected in how the estimate is reported and used in later stages.
- **Developers should consider a range of deployment options.** Staged release, structured access for vetted researchers and auditors, and controlled download for approved users preserve many of the benefits of openness, such as external scrutiny and wider access to capable models, without incurring the full risk of unrestricted weight release [51, 52]. The relevant question is often which release channel is appropriate rather than whether to release at all. The GLM-5.3 release illustrates this in practice: citing dual-use risks, access was structured so that vetted security partners evaluated the model in controlled settings first, with broader access and API availability following, and full weights were published only once safety evaluations and release preparations were complete [53]. Ultimately, developers should take context into account and weigh the added risk of open release against the added benefit, since different actors will reach different conclusions about the same model based on their risk tolerance, ecosystem, and the specific release mechanism under consideration [24].

These considerations apply to a single release, but small increases in risk across successive releases can compound over time into substantial increases in total risk, even when each release looks acceptable on its own [3]. Evaluation practices should therefore be designed specifically for open-weight releases [11], and developers should periodically review whether cumulative capability trends have shifted the appropriate threshold for release.

The risk treatment strategy in each zone should be adapted for the open-weight case:

| Zone Classification | Decision & Treatment Strategy |
|---|---|
| **Green Zone** (Broadly Acceptable) | **Full open-weight release**<br>● Residual risk remains broadly acceptable for both the base model and potential post-release modifications (e.g., uncensored or maliciously fine-tuned variants).<br>● Proceed with full open-weight release. Requires baseline safety practices, model documentation, and post-release impact monitoring via available feedback channels. |
| **Yellow Zone** (Tolerable) | **Restricted and/or staged release**<br>● Risks are tolerable only when a set of mitigation measures are implemented upstream by open-weight developers (throughout the model development lifecycle) and by hosting platforms.<br>● Release proceeds under one of a spectrum of alternatives, including staged or gradual release, structured access for vetted researchers and external auditors, or gated access for specific use cases or hosting via API while weight release is evaluated. |
| **Red Zone** (Unacceptable) | **Suspension of release**<br>● Release is suspended or postponed whenever a model crosses domain-specific red lines[1] pending re-evaluation.<br>● If an open-weight model is reclassified as Red Zone after release, mitigation must shift to ecosystem-level responses, as open-weight release cannot be reversed. |

Risk evaluation decisions are difficult to reverse once weights are released. As these models can be modified arbitrarily, used without oversight, and spread irreversibly, this stage shapes the risk mitigation and risk governance stages that follow [6]. Risk mitigation for open-weight models must concentrate on measures that take effect before release or continue to operate once weights are distributed, since in-use controls at the deployment layer are not available. Risk governance must extend beyond the original developer to the ecosystem of downstream variants that any open release produces.

[1] See section 2.2. for the red line specifications for cyber offense risks, biological risks, chemical safety risks, large-scale persuasion and harmful manipulation risks, and loss of control risks in *Framework 2.0*.

# Risk Mitigation

This section maps mitigation measures for open-weight AI onto the model lifecycle, from pre-development through post-deployment. This mirrors the structure of recent taxonomies [34, 54, 55], which converge on the view that risk mitigation must occur at multiple points. The section is organized around a central difference between open-weight and closed-weight models: because open-weight release removes most of the controls that typically operate at the deployment layer, the mitigation burden moves upstream to pre-training and outward to the broader ecosystem.

This is where this framework departs most sharply from the *Framework 2.0* [1]. In that document, the yellow zone risk mitigation toolkit relies on user KYC mechanisms, content input and output restrictions on APIs, and oversight of where and how models are deployed. All three techniques depend on the developer retaining control of the interface between users and the model, which is precisely what is lost with open-weight release.

## Pre-development stage

Pre-development mitigations act before the weights exist, which makes them uniquely durable: a capability that was never learned cannot be reactivated by a downstream user. Training data curation is the leading candidate here, and recent evidence suggests it is one of the strongest layers of defense available for open-weight release.

- **Filtering hazardous content out of training data appears to be one of the most durable defenses currently available.** Recent studies show that models pretrained on filtered data resist adversarial fine-tuning far more strongly than models patched after training, with no cost to performance on unrelated tasks [56, 57, 58].
- **Data curation is not a complete defense.** When hazardous information is supplied at inference time through retrieval or tool augmentation, data curation limits what the filtered model knows but does not prevent it from using dangerous information that is fed in [57]. Research has also not yet established which training content leads to which capabilities, or how to scale reliable methods for identifying what to filter [54]. Due to these limits, risk mitigation should employ defense in depth, rather than relying on data curation alone.

## Development stage

Once pre-training is complete, risk mitigation shifts to techniques that are applied to the trained model. For open-weight models, any such safeguard must survive adversarial fine-tuning to remain robust once weights are public. Due to this constraint, most research in this area focuses on tamper-resistant fine-tuning, alongside a set of emerging methods that aim to embed safety more deeply in the model or its runtime environment.

- **Tamper-resistant fine-tuning.** A growing family of mitigation techniques aim to embed safeguards more deeply in the model, so that removing them requires efforts more comparable to full retraining than to a few steps of fine-tuning [43, 59, 60]. The *International AI Safety Report 2026* characterizes this line of work as promising but (so far) challenging to achieve robustly [3]. In particular, “unlearning” techniques have been shown to obfuscate hazardous knowledge rather than remove it, and state-of-the-art methods have been broken by fine-tuning models on as few as ten unrelated examples or by editing specific directions in activation space [61]. Even successful unlearning appears insufficient in models with strong in-context learning, since removed knowledge can be reintroduced through prompts [62]. More recent work attempts to separate selected capabilities into modules during pre-training that can be withheld at release, though this has so far been demonstrated only at small scale [63].

- **Scaffolding-based methods.** These are add-on safeguards, such as guard models that screen prompts for malicious intent and check responses for safety violations before they reach the user. Anyone with the weights can trivially disable or bypass them, but they still help prevent accidental misuse, deter unsophisticated actors, and give downstream developers a starting point for their own safety systems [54].
- **Cryptographic and hardware-based methods.** Nascent approaches such as parameter-level encryption and GPU-side key management would let a model be modified or run only when an authorized decryption step is completed [64]. Recent safety research explores verifiable mathematical security boundaries for AI models [65]. These directions are conceptually promising for open-weight risk mitigations but have not yet been demonstrated at frontier model scale.

## Deployment stage

Deployment is itself a risk mitigation point. Because open-weight release cannot be revoked, two choices at this stage matter especially: which release channel to use, and what documentation to release alongside the weights.

- **Release strategy sits on a spectrum.** Deployment options range from fully closed API access to fully open weights, with intermediate points such as gated or researcher-only access [51]. Staged release involves opening access gradually and lets developers and defenders monitor uptake at each step [16, 54, 66].
- **Documentation released alongside weights enables independent scrutiny.** Detailed model cards, technical reports, and evaluation data give external researchers the context needed to study a system and its downstream effects [54]. Examples of best practices include releasing a safety classifier alongside a technical report, benchmarks, and weights [67], as well as peer review [44, 68].

## Post-deployment stage

Once weights are released, focus shifts to monitoring and mitigating the risks in an ecosystem of downstream model variants:

- **Model provenance techniques.** Methods include recovering a model's heritage by inferring which base model a variant descends from [69] and watermarking model weights so that copies can be identified even after modification [70]. Both can be defeated with some effort, but they raise the cost of anonymous misuse and support post-hoc attribution [54]. Complementary work proposes charting the open-model ecosystem itself, mapping public repositories into a searchable atlas that supports downstream applications, such as attribute prediction and detecting derived models [71].

- **Data provenance techniques trace AI-generated content back to its source.** Imperceptible watermarks and metadata standards let researchers study the spread of AI-generated content, including from open-weight models. These can also be removed, but they are already useful in practice and are gradually being adopted across the ecosystem [54].
- **Distribution platforms become the natural point for ecosystem-level access controls.** Because the developer no longer mediates access, know-your-customer measures and usage monitoring shift outward to distribution platforms and downstream deployers, which are the main remaining levers [54].

# Risk Governance

The primary objective of the risk governance stage is to establish organizational structures, oversight mechanisms, and accountability frameworks that ensure the entire risk management process is rigorously implemented, continuously monitored, and regularly adapted to evolving threats and capabilities. International risk management standards emphasize that regardless of the assessment methods used, human and organizational factors must be given due recognition when identifying and describing risks [22]. Once weights are released, the organizational unit responsible for governance is no longer a single developer, but a distributed set of actors along the value chain. China's National Technical Committee 260 on Cybersecurity (TC260) *AI Safety Governance Framework 2.0* addresses this directly: it argues that the open-source ecosystem should have stronger safety measures, calling model providers and open-source communities to develop shared rules, to inform users of the risks and security responsibilities that come with a downloaded model, and to specify prohibited uses of open-weight models [5]. The Partnership on AI reaches a similar conclusion, listing model providers, adapters, hosting services, and downstream application developers as relevant parties, each of whom occupies a distinct position with different levers available [34].

## Internal Governance

The internal governance section in *Framework 2.0* concentrates on the developer's own processes: risk committees, review boards, and internal accountability lines [1]. For open-weight releases, these mechanisms remain necessary, but they are no longer sufficient, since the developer's internal controls do not reach the actors who fine-tune, host, or redeploy the model after release.

- **Developer-side internal governance remains the foundation.** Regulation such as the EU AI Act's Code of Practice requires providers of open-license GPAI models with systemic risk to conduct model evaluations, run adversarial testing, track and report serious incidents, and maintain cybersecurity protections [72]. Industry-led guidance emphasizes documenting development processes, sharing evaluation frameworks, and releasing openly licensed artifacts that others can build on [73], as well as using internal audits and documentation to monitor failures, performance drift, and unsafe behavior [74]. Other proposals recommend that developers establish open source governance committees, conduct pre-release security risk assessments, and prepare emergency response plans within their own organizations [75].
- **Equivalent internal processes are needed further down the value chain.** Model adapters, hosting services, and downstream deployers each need their own governance structures adapted to their role, since developer-side controls do not travel with the weights [34]. The EU AI Act's Code of Practice operationalizes the boundary between developer and downstream provider thus: a downstream modifier becomes the "provider" of a general-purpose AI model once modifications significantly change its generality, capabilities, or systemic risk [76].
- **Uncertainty over accountability remains a governance gap.** Even with distributed internal governance in place, it is not yet settled who bears responsibility when open-weight models are modified for harmful purposes after release [3]. This framework treats this as an open question, and the transparency and external oversight mechanisms in the following subsection are partly designed to address this.

## Transparency and Social Oversight

Because open weights are public, external actors can inspect, evaluate, and identify vulnerabilities in ways that closed release does not allow. This makes transparency a governance mechanism in its own right.

- **Open weights let outside actors do work that closed release cannot support.** Some scholarship has argued that openness is not merely compatible with AI safety but conducive to it: bias and security flaws surface and are patched sooner, evaluations gain credibility through review across the developer community, and open code provides verifiable evidence for external auditing [77, 78]. This upside, however, is not automatic, and depends on the ecosystem having the skills and coordination to act on findings.
- **Monitoring and disclosure obligations need to extend to frontier models' internal reasoning processes.** Frontier models are increasingly relying on extended internal reasoning and deliberation that are not fully observable in final outputs, and that can be sustained across long, multi-turn agentic interactions. This creates greater scope for deceptive alignment—scenarios where a model internally formulates a harmful intent while its outward-facing outputs remain compliant [79, 80]. Model and system cards should therefore disclose whether extended internal reasoning is enabled, the type of safety-monitoring mechanism applied (i.e., whether the full reasoning chain is inspected, or just the final output), and the residual risk that monitoring does not address.
- **Model weight and inference hosting platforms are where oversight actually happens.** Platforms such as HuggingFace, GitHub, and ModelScope are the places where model cards, technical reports, and evaluation data are searchable and downloadable, and where researchers, auditors, and civil society can inspect a system after release [81]. Platform oversight, however, has limits: use restrictions for original models are not reliably preserved in derivatives re-uploaded to the platforms [38]. Complementary practices at the developer end, including full auditor access before release and staged release to progressively larger groups, can help surface flaws before wide availability [3].